\documentclass[aps,prc,groupedaddress,twocolumn,showpacs,amsmath,amssymb,floatfix]{revtex4}
\usepackage{graphics}% Include figure files
\usepackage{dcolumn}% Align table columns on decimal point
\usepackage{longtable}

\begin{document}
\title{Deuteron-induced reactions on manganese at low energies}

\author{M.~Avrigeanu$^{1}$\footnote{marilena.avrigeanu@nipne.ro},
E.~\v Sime\v ckov\'a$^{2}$\footnote{simeckova@ujf.cas.cz},
U.~Fischer$^3$,
J.~Mr\'azek$^2$,
J.~Novak$^2$,
M.~\v Stef\'anik$^2$,
C.~Costache$^1$, 
V.~Avrigeanu$^1$
}

\affiliation{$^1$Horia Hulubei National Institute for R and D in Physics and Nuclear Engineering, P.O. Box MG-6, 077125 Bucharest-Magurele, Romania}
\affiliation{$^2$Nuclear Physics Institute CAS, 25068 \v Re\v z, Czech Republic}
\affiliation{$^3$Euratom/FZK Fusion Association, Karlsruhe Institute of Technology (KIT), Hermann-von-Helmholtz-Platz, 1, 76344 Eggenstein-Leopoldshafen, Germany}

%\date{\today}

\begin{abstract}
\noindent
{\bf Background:}
The scarce data systematics and complexity of deuteron interactions demand the update of both the experimental database and theoretical frame of deuteron activation cross sections. Various reactions induced by neutrons and protons following the deuteron breakup (BU) should be also taken into account. On the other hand, deuteron reaction cross sections recommended recently for high-priority elements are still based on data fit without predictive power.

\smallskip

\noindent
{\bf Purpose:}
Accurate new measurements of low-energy deuteron-induced reaction cross sections for monoisotopic ($^{55}$Mn) natural manganese target enhance the related database as well as the opportunity of an unitary and consistent account of the related reaction mechanisms. 

\smallskip

\noindent
{\bf Methods:}
Activation cross sections of $^{54,56}$Mn, and $^{51}$Cr nuclei by deuterons  on $^{55}$Mn were measured at energies $\leq$20 MeV by the stacked-foil technique and high resolution gamma spectrometry at the U-120M cyclotron of CANAM, NPI CAS. Then all available data for deuterons on $^{55}$Mn up to 50 MeV are analyzed 
paying particular attention to BU and direct reaction (DR) mechanisms. 

\smallskip

\noindent
{\bf Results:}
Newly measured activation cross sections strengthen the deuteron database at low energies, at once with a consistent account for the first time of all available data. 

\smallskip

\noindent
{\bf Conclusions:}
Due account of deuteron-induced reactions on $^{55}$Mn, including particularly the new experimental data at low energies, is provided by a suitable BU and DR assessment.
\end{abstract}


\maketitle

\section{Introduction}
\label{Sec1}

Ongoing fusion research programs (ITER, IFMIF, SPIRAL2-NFS) \cite{iter} and medical investigations using accelerated deuterons triggered even a decade ago an update of the experimental  database and theoretical account of deuteron-induced reactions. The related FENDL-library project \cite{FENDL} was motivated essentially by the scarce deuteron data systematics and complexity of deuteron interactions. Various reactions induced by neutrons and protons following the deuteron breakup (BU) should be also taken into account.

More recently, full parametrization of the deuteron monitor reactions and therapeutic radionuclides-production cross sections have been recommended by Hermanne {\it et al.} \cite{hermanne}, and Engle {\it et al.} \cite{engle}. 
Thus, beyond the FENDL \cite{FENDL} concern of  improved theoretical analysis, fit of the available data by a least-squares method with Pad\'e approximations of variable order has been involved to evaluate the deuteron-induced production cross sections and corresponding uncertainties.
Actually, deficiencies in describing the elastic and especially the inelastic BU that dominate the deuteron reaction cross section at beam energies comparable with the Coulomb barrier motivated the recommended Pad\'e fits \cite{engle}.
Consequently, the present work aims to strengthen the database of deuteron-induced reactions up to 20 MeV on $^{55}$Mn target nucleus, as well as a deeper understanding of the involved reaction mechanisms. The weak points of an evaluation procedure that may lead to disagreement with experimental data are also carefully concerned.

The particular choice of the Mn element for this work is related to the design of the structural components of accelerators for which the deuteron-induced activation has a major importance for material damage and radioactivity risks. 
Thus, there is a 2\% Mn presence in the SS-316L steel alloys that are considered as option for the beam guides of the DONES project \cite{dones}. 
It is therefore of interest to complete the recent analyses of deuteron interaction with V, Cr, Fe, Co, Ni, and Cu \cite{VCo,Crd,Fed,Nid,Cud} looking also for the consistent account of BU and direct reaction (DR) contributions to activation cross sections. This is as well essential in view of significant disagreement between recent measurements and calculated data \cite{ditroi,tark}.

The experimental setup as well the new measured data are described in Sec.~\ref{exp}. 
Among the models involved in this work is firstly an energy-dependent optical potential for deuterons on $^{55}$Mn in Sec.~\ref{omp}. 
The BU analysis is described in Sec.~\ref{DI} as well as DR account using the computer code FRESCO \cite{FRESCO}, while the main points of the pre-equilibrium emission (PE) and compound nucleus (CN) contributions of the code TALYS-1.9 \cite{talys} are given in Sec.~\ref{PE+CN}. Discussion of the measured and calculated cross sections is in Sec.~\ref{Activation} makes reference to the newest TENDL-2017 evaluated data library \cite{TENDL} as well. Finally, conclusions of this work are given in Sec.~\ref{Sum}.

\section{Measurements} \label{exp}

The irradiation was carried out on infrastructure of the Center of Accelerators and Nuclear Analytical Methods CANAM \cite{CANAM} of the Nuclear Physics Institute of the Czech Academy of Sciences (NPI CAS). 
Deuterons were accelerated by the variable-energy cyclotron U--120M.
D$^+$ beam was extracted using a stripping foil and the ion beam was delivered to the reaction chamber.
The mean beam energy was determined with an accuracy of 1\%, with full width at half maximum (FWHM) of 1.8\%.

The collimated deuteron beam impinged the stack of foils placed in a cooled reaction chamber that served also as a Faraday cup. The measured MnNi foils (Goodfellow product, 25 $\mu$m declared thickness) containing  88\% monoisotopic ($^{55}$Mn) natural manganese and 12\% Ni were interleaved by monitoring/degrading Al foils (Goodfellow product - 99.99\% purity, 50 $\mu$m declared thickness). 

The irradiation was carried up in two runs to check an internal consistency of the measurement. The characteristics of the single runs are given in Table~\ref{tab:runs}.
The mean energy, the energy thickness and the energy spread in each foil were simulated by the SRIM 2008 package \cite{SRIM}. 

\begin{table} [b]
\caption{\label{tab:runs} Characteristics of single runs. }
\begin{tabular}{ccccc}\\
\hline \hline
Run & Initial &  Total & Irradiation &  Mean  \\	
no. &  energy &  charge &   time     & current\\	
    &  (MeV)  &  ($\mu$C) &  (s)     &($\mu$A)\\ 
\hline
1  &  19.76  &     96.05 &  303      & 0.317  \\ 
2  &  19.5  &     736.87 & 1833      & 0.402  \\	
\hline \hline
\end{tabular}
\end{table}

\begin{table} [t]
\caption{\label{tab:decay} Half-lives, main $\gamma$ lines, and their intensities \cite{chu99} of the isotopes observed from irradiated $^{55}$Mn.}
\begin{tabular}{cccc}\\
\hline \hline
    Isotope   & T$_{1/2}$ & E$_{\gamma}$(keV) & I$_{\gamma}$(\%) \\
\hline
$^{56}$Mn     & 2.5785 h   & 846.77           &98.9 \\
              &            & 1810.77          & 27.2 \\
$^{54}$Mn     & 312.3 day  & 834.25           & 99.98 \\
$^{51}$Cr     & 27.7025 day& 320.08           & 10    \\
\hline \hline
\end{tabular}
\end{table}
 
\begingroup
\squeezetable
\begin{table} [!b]
\caption{\label{tab:cs} Measured reaction cross sections (mb) for deuterons incident on natural manganese ($^{55}$Mn). The uncertainties are given in parentheses, in units of the last digit.}
\begin{tabular}{cccc}\\
\hline \hline
 Energy  & \multicolumn{3}{c}{Reaction}\\
           \cline{2-4}
   (MeV) & \rotatebox{00}{Mn$(d,p)^{56}$Mn}
	       & \rotatebox{00}{Mn$(d,x)^{54}$Mn}
				 & \rotatebox{00}{Mn$(d,x)^{51}$Cr}\\
\hline
19.49 (27)& 87.7  (61)                           \\
19.22 (28)& 75.2  (47)& 133.1   (77) & 0.353 (44)\\
19.04 (26)& 90.4  (57)& 108.0   (69)             \\
18.20 (28)& 84.4  (52)&  89.0   (52) & 0.119 (40)\\
17.53 (28)& 94.1  (76)&  50.8   (36)             \\
17.13 (29)& 90.9  (55)&  52.7   (30)             \\
16.50 (30)&115.9  (67)                           \\
16.02 (30)&100.1  (71)&  30.49 (176)             \\
15.90 (32)&111.3  (71)&  17.89 (124)             \\
14.85 (32)&111.4  (75)&  20.32 (118)             \\
14.81 (32)&131.3  (86)                           \\
14.12 (35)&148.6 (175)&  16.25  (99)             \\
13.59 (35)&125.5 (101)&  16.04  (92)             \\
12.22 (39)&145.6 (114)&  13.37  (78)             \\
12.17 (36)&165.5 (146)&  11.82  (68)             \\
10.73 (42)&169.2 (133)&  10.58  (61)             \\
 9.93 (45)&151.3 (312)&   7.74  (55)             \\
 9.08 (47)&199.4 (155)&   6.84  (39)             \\
 7.17 (55)&231.2 (173)&   2.47  (15)             \\
 7.16 (55)&256.5 (239)&   1.69  (18)             \\
 4.80 (72)&229.5 (137)                           \\
 3.13 (97)&134.4 (132)                           \\
\hline \hline
\end{tabular}
\end{table}
\endgroup

The $\gamma$ rays from the irradiated foils were measured repeatedly by two calibrated high-purity germanium (HPGe) detectors of 50\% efficiency and FWHM of 1.8 keV at 1.3 MeV. Experimental reaction rates were calculated from the specific activities at the end of the irradiation and corrected for the decay during the irradiation using the charge measurement and MnNi foil characteristics as well. The measurement with different cooling times lasted up to 100 days after irradiation. The decay data of the isotopes observed from irradiated MnNi foils \cite{chu99} are given in Table~\ref{tab:decay}.

The experimental cross sections of the $^{55}$Mn$(d,p)^{56}$Mn, $^{55}$Mn$(d,x)^{54}$Mn and $^{55}$Mn$(d,x)^{51}$Cr reactions are given in Table~\ref{tab:cs}.
The energy errors take into account the energy thickness of each foil and the initial-energy spread error. Cross-section errors are composed of statistical errors in activity determination and systematical errors of charge measurement uncertainty ($\sim$5\%), foil thickness uncertainty (2\%) and uncertainty of HPGe detector efficiency determination (2\%). 
The measured cross sections are in a good agreement with recent data and will be discussed in Sec.~\ref{Activation}.

\section{Nuclear model analysis}
\label{models}

\subsection{Optical potential assessment}
\label{omp}

The optical model potential (OMP) parameters obtained and/or validated by analysis of deuteron elastic-scattering angular distributions are then used within calculation of all deuteron-induced reaction cross sections. 
Consequently, the simultaneous analysis of elastic-scattering and activation data is essential for nuclear model calculations using a consistent input parameter set \cite{VCo,Crd,Fed,Nid,Cud,Ald,Nbd}. 
			
\begin{figure} [h]
\resizebox{0.80\columnwidth}{!}{\includegraphics{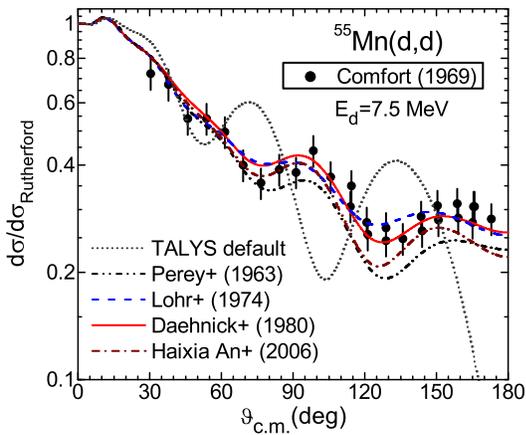}}
\caption{(Color online) Comparison of the measured \cite{comfort} and calculated deuteron elastic-scattering angular distributions using, beyond the TALYS default double-folding approach \cite{talys} (dotted curve), the deuteron global OMPs of Perey-Perey \cite{PP} (dash-dot-dotted curve), Lohr-Haeberli \cite{LH} (dashed curve), Daehnick et al. \cite{dah} (solid curve), and Haixia -Chonghai \cite{haix} (dash-dotted curve) for $^{55}$Mn target nucleus. }
\label{55Mnd_elasticf}
\end{figure}

However, the OMP analysis in this case makes use of only one measurement for deuteron elastic-scattering angular distribution on $^{55}$Mn at 7.5 MeV incident energy \cite{comfort}.  
The best description of the measured data at a particularly low incident energy has been obtained using the deuteron OMP of Daehnick {\it et al.} \cite{dah} versus TALYS default double-folding approach \cite{talys} as well as predictions given by Perey-Perey \cite{PP}, Lohr-Haeberli \cite{LH}, and An-Cai \cite{haix} global optical potentials  (Fig.~\ref{55Mnd_elasticf}). This potential was furthermore involved in the rest of this work.
	
\subsection{Direct Interactions}
\label{DI}

The specific BU and DR noncompound processes make deuteron-induced reactions substantially different from reactions with other incident particles. 
Actually, noticeable discrepancies between the measured activation data and theoretical model results follow mainly the disregard of these direct interactions (DIs) peculiarity \cite{VCo,Crd,Fed,Nid,Cud,Ald,Nbd}. 

\begin{figure*} %[h]
\resizebox{1.6\columnwidth}{!}{\includegraphics{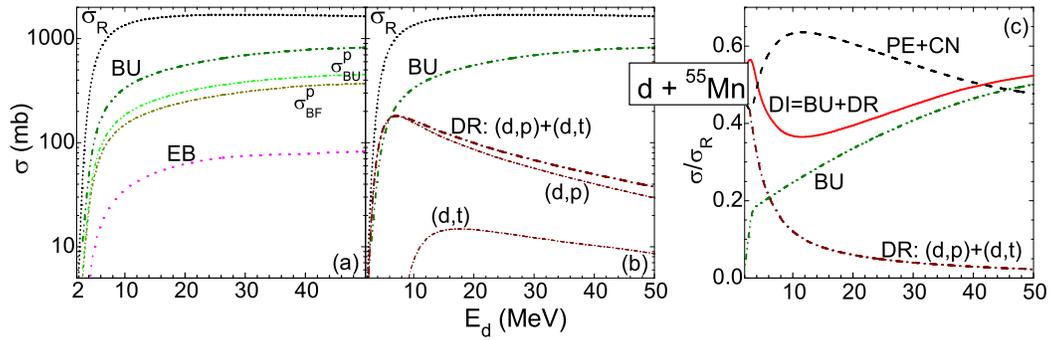}}
\caption{(Color online) The cross-section energy dependence of (a,b) deuteron total-reaction (short-dotted curves) and BU (dash-dot-dotted curves), (a) proton-emission BU (thin dash-dot-dotted curve) and BF (short dash-dotted), EB (dotted), (b) DR (dash-dotted), stripping $(d,p)$ (thin dash-dotted) and pick-up $(d,t)$ (short dash-dotted), and (c) the $\sigma_R$ fractions of BU (dash-dot-dotted), DR (dash-dotted), DI (solid), and PE+CN cross sections of deuteron interactions with $^{55}$Mn (see text).}
\label{55Mnd_DIf}
\end{figure*}

\subsubsection{Breakup}
\label{BU}

There are two distinct BU processes in the Coulomb and nuclear fields of the target nucleus. The former is the elastic breakup (EB) in which the target nucleus remains in its ground state and none of the deuteron nucleons interacts with it. The later is the inelastic breakup or breakup fusion (BF), where one of these deuteron constituents interacts nonelastically with the target nucleus. 

Our description of the deuteron breakup mechanism is based on the parametrizations \cite{breakup0,breakup} for (i) the cross sections $\sigma_{BU}^{p}$  of total BU proton emission, and (ii) the EB cross sections $\sigma_{EB}$. Equal inelastic-breakup cross sections $\sigma_{BF}^{p}$ and $\sigma_{BF}^{n}$ have been assumed for the breakup neutron and proton, so that $\sigma_{BU}^{n/p}$=$\sigma_{EB}$+$\sigma_{BF}^{n/p}$ while the total breakup cross section is $\sigma_{BU}$=$\sigma_{EB}$+2$\sigma_{BF}^{n/p}$  \cite{breakup,Pad}.
Actually our parametrizations have concerned the total BU nucleon-emission and EB fractions of the deuteron OMP total-reaction cross section $\sigma_{R}$, i.e. $f_{BU}^{n/p}$ = $\sigma^{n/p}_{BU}/\sigma_R$ and  $f_{EB}$=$\sigma_{EB}/\sigma_R$, respectively. 
The dependence of these fractions on the deuteron incident energy $E$ and target-nucleus atomic $Z$ and mass $A$ numbers was obtained through analysis of the experimental systematics of deuteron-induced reactions on target nuclei from $^{27}$Al to $^{232}$Th and incident energies up to 80 MeV for the former:
\begin{eqnarray}\label{eq:01}
f^{n/p}_{BU}=0.087-0.0066 Z + 0.00163 ZA^{1/3} +  \:  \nonumber \\
0.0017A^{1/3}E - 0.000002 ZE^2 \:\:\:\:  ,
\end{eqnarray}
but within a more restricted energy range up to 30 MeV for the later \cite{pamp78_all}:
\begin{eqnarray}\label{eq:02}
f_{EB}=0.031-0.0028 Z + 0.00051 ZA^{1/3} +  \:  \nonumber \\
0.0005A^{1/3}E - 0.000001 ZE^2 \:\:\:\:  
\end{eqnarray}
More details are given elsewhere \cite{breakup0,breakup,Pad}. 

The comparison of the measured $\sigma^p_{BU}$ with microscopic breakup cross sections \cite{neoh,carlson} as well as the empirical parametrization pointed out the correctness of our breakup analysis \cite{breakup,varenna18}. However, the improvement of the deuteron breakup description requires, beyond the increase of its own data basis, also complementary measurements of $(d,px)$ and $(n,x)$, as well as $(d,nx)$ and $(p,x)$ reaction cross sections for the same target nucleus, within the related incident-energy ranges. 

The energy dependence of $\sigma_R$, $\sigma_{BU}$, $\sigma_{BU}^{p}$, as well as the BU components $\sigma_{EB}$ and $\sigma_{BF}^{p}$ for deuteron interactions with $^{55}$Mn target nucleus are compared in Fig~\ref{55Mnd_DIf}(a). 
The breakup excitation functions increase with deuteron-energy increasing, including its dominant BF component that is quite important for the analysis of the following two opposite BU effects on deuteron activation cross sections. 

Firstly, the leakage of the initial deuteron flux toward the breakup process reduces the total-reaction cross section that should be shared among different outgoing channels by a reduction factor $(1-\sigma_{BU}/\sigma_R)$. 

On the other hand, the BF component brings contributions to different reaction channels of deuterons interactions with the target nuclei \cite{VCo,Crd,Fed,Nid,Cud,Ald,Nbd,Pad}. Thus, the absorbed proton or neutron following the deuteron breakup contributes to the enhancement of the corresponding $(d,xn)$ or $(d,xp)$ reaction cross sections, respectively. The compound nuclei in reactions induced by the BF nucleons differ by one unit of the atomic mass and maybe of also the atomic number in comparison with deuteron-induced reactions, the partition of the BF cross section among various residual-nuclei population being triggered by the energy spectra of the breakup nucleons and the excitation functions of the reactions induced by these nucleons on the target nuclei. 

In order to calculate the BF enhancement of, e.g., the $(d,xn)$ reaction cross sections, the BF proton-emission cross section $\sigma^{p}_{BF}$ should be (i) multiplied by the ratios $\sigma_{(p,x)}$/$\sigma^p_R$, corresponding to the enhancing reaction, (ii) convoluted with the Gaussian line shape distribution of the BF--proton energy $E_p$ for a given deuteron incident energy $E_d$, and (iii) integrated over the BF proton energy. Consequently, the BF--enhancement cross section has the form \cite{Fed,Nid,varenna18}:
\begin{eqnarray}\label{eq:1}
\sigma^{p,x}_{BF}(E_d) = \sigma^{p}_{BF}(E_d) \int dE_p \: \: \frac{\sigma_{(p,x)}(E_p)} {\sigma^p_R} \: \nonumber \\
\frac {1} {(2\pi)^\frac{1}{2} w} exp[ - \frac{(E_{p}-E^0_{p}(E_d))^2}{2w^2}]\:\: ,
\end{eqnarray}
\noindent
where $B_{d}$ is the deuteron binding energy, $\sigma^p_R$ is the proton total-reaction cross section, $x$ stands for various $\gamma$, $n$, $d$, or $\alpha$ outgoing channels, while the Gaussian distribution parameters $w$ and $E^0_{p}$ given by Kalbach \cite{kalb03} were used. 
Interpolation of experimental nucleon-induced reaction cross sections (e.g., $\sigma_{(p,x)}$) from the EXFOR library \cite{EXFOR} or from newest TENDL library has been involved within estimation of the BU enhancement \cite{VCo,Crd,Fed,Nid,Cud,Ald,Nbd,Pad}, in order to reduce as much as possible the supplementary uncertainties brought by additional theoretical calculations. The nucleon total-reaction cross sections $\sigma^{n/p}_R$ have been obtained by using the same optical potentials that have been involved within the stripping and statistical nucleon emission calculations.

On the whole, the enhancing effect of the breakup mechanism is important mainly for describing the excitation functions for second and third chance emitted-particle channels.
The BF enhancements due to the BU protons and neutrons emitted during the deuteron interaction with $^{55}$Mn, through the $(p,x)$, and $(n,x)$ reactions populating various residual nuclei, are discussed in Sec.~\ref{Activation}.

\begin{figure*}
\resizebox{2.06\columnwidth}{!}{\includegraphics{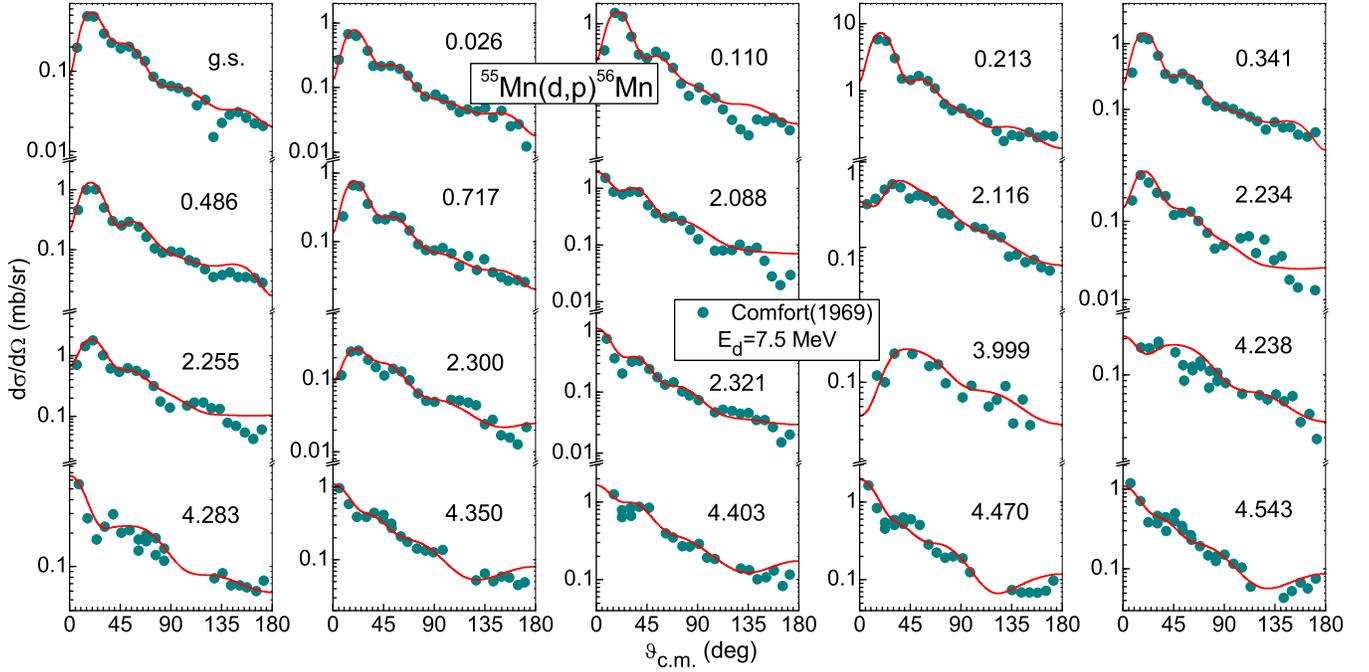}}
\caption{(Color online) Comparison of measured (solid circles) \cite{comfort} and calculated (solid curves) proton angular distributions of $^{55}$Mn$(d,p)^{56}$Mn stripping transitions to states with excitation energies in MeV, at the incident energy of 7.5 MeV.}
\label{55Mn_dp}
\end{figure*}
\begin{figure*}
\resizebox{2.06\columnwidth}{!}{\includegraphics{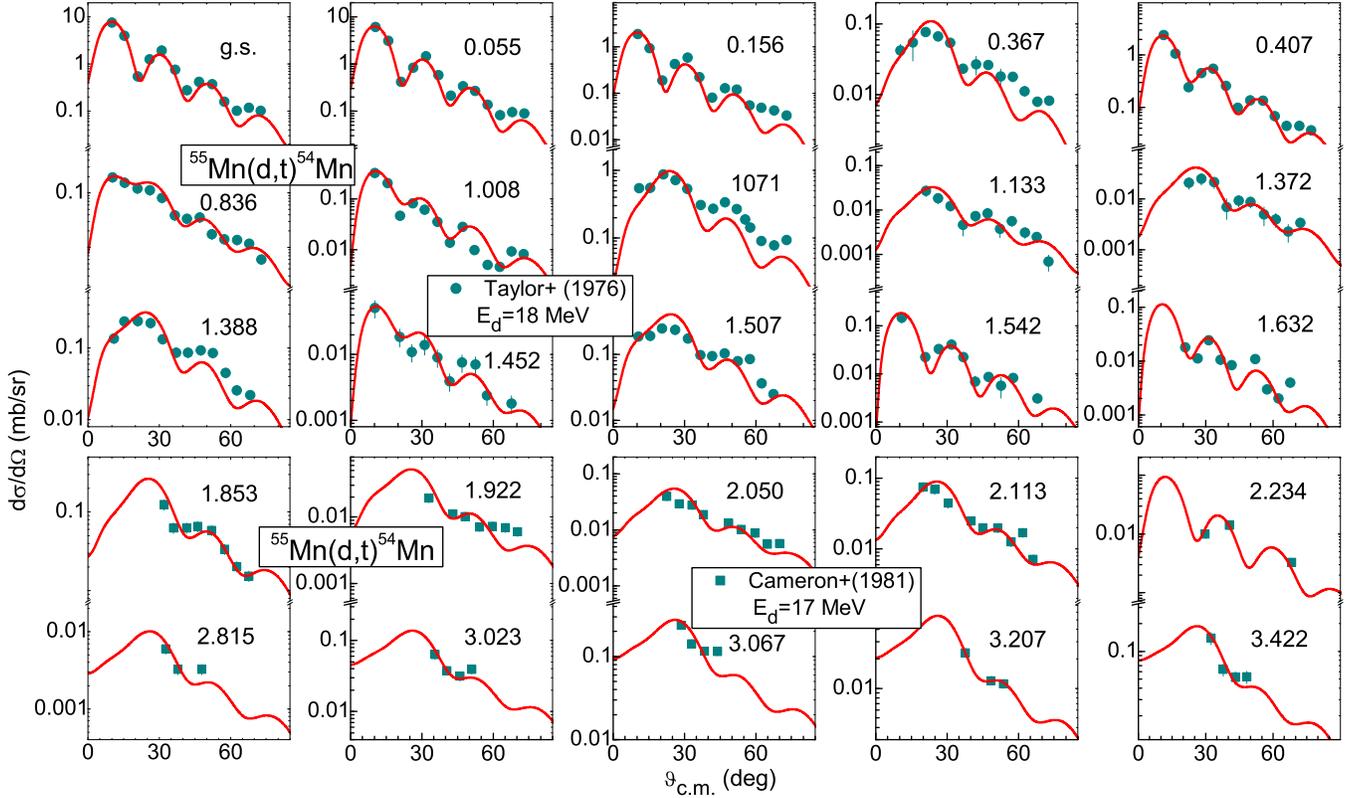}}
\caption{(Color online) Comparison of measured (solid circles) \cite{taylor,cameron} and calculated (solid curves) triton angular distributions of $^{55}$Mn$(d,t)^{54}$Mn pick-up transitions to states with excitation energies in MeV, at the incident energies of 18 and 17 MeV.}
\label{55Mn_dt}
\end{figure*}

\subsubsection{Direct reactions}
\label{DR}

In addition to the BU assessment, an accurate estimation of the DR cross sections is necessary to obtain finally the correct PE+CN population of various residual nuclei. However, poor attention was given so far to this issue in deuteron activation analysis also because the DR cross-section account is conditioned by the available experimental spectroscopic factors, outgoing particle angular distributions, or at least the differential cross-section maximum values. 
On the other hand, for nuclear reactions involving projectiles and ejectiles with different particle numbers, TALYS includes additionaly to PE only the account of the continuum stripping, pick-up, breakup and knock-out reactions using Kalbach's phenomenological contribution for these mechanisms \cite{talys}.

The appropriate calculation of the DR stripping and pick-up mechanism contributions within this work used the distorted-wave Born approximation (DWBA) formalism within the code FRESCO \cite{FRESCO}. 
The post/prior form distorted-wave transition amplitudes for stripping and respectively pick-up reactions, and the finite-range interaction have been considered. 
The $n$-$p$ effective interaction in deuteron \cite{kamimura86} as well as $d$-$n$ effective interaction in triton \cite{triton} were assumed to have a Gaussian shape, while the transferred-nucleon bound states were generated in a Woods-Saxon real potential \cite{VCo,Crd,Fed,Nid,Cud,Ald,Nbd,Pad}. 
The populated discrete levels and the corresponding spectroscopic factors available within the ENSDF library \cite{BNL} were used as starting input for the DWBA calculations of the $(d,p)$ stripping \cite{NDS_dp}, and $(d,t)$ pick-up \cite{NDS_dt} excitation functions. 

Experimental angular distributions of particle emission in deuteron-induced DRs on $^{55}$Mn there are only for the $(d,p)$ stripping \cite{comfort} and $(d,t)$ pick-up \cite{taylor,cameron} reactions. 
A detailed analysis of the former reaction at 7.5 MeV incident energy has made possible the calculation of almost total $(d,p)$ stripping contribution to the population of $^{56}$Mn residual nucleus nucleus, of 129 discrete levels up to the excitation energy of  $\sim$ 5.6 MeV \cite{comfort,NDS_dp}.
On the other hand, the pick-up $(d,t)$ cross-section calculations concerned the analysis of the triton angular distributions measured at incident energies of 17 \cite{taylor} and 18 \cite{cameron} MeV for 30 discrete levels up to $\sim$ 3.7 MeV \cite{taylor,cameron,NDS_dt}. 
The suitable description of the measured proton (Fig.~\ref{55Mn_dp}) and triton (Fig.~\ref{55Mn_dt}) angular distributions, as well as of their maximum values, has validated the correctness of the $(d,p)$ and $(d,t)$ excitation-function calculations shown in Fig.~\ref{55Mnd_DIf}(b). 

However, due to the missing data on $(d,n)$ stripping and $(d,\alpha)$ pick-up reactions, the sum $\sigma_{(d,p)}$ + $\sigma_{(d,t)}$ could stand only as a lower limit of the DR component shown also in Fig.~\ref{55Mnd_DIf}(b). 
It has a significant maximum around $E_d\sim$7 MeV mainly due to the $(d,p)$ strong stripping processes. 
Then, it results a slow decrease at higher deuteron energies while $\sigma_R$ reaches its maximum value and remains almost constant for $E_d$$>$ 25 MeV. 
The major role of the stripping $(d,p)$ reactions to the activation cross sections of $^{56}$Mn residual nucleus, for deuteron interaction with $^{55}$Mn, will be shown in Sec.~\ref{Activation}. The exclusive contribution of $(d,t)$ pick-up process at the lowest incident energies, up to the thresholds of the $(d,nd)$ and $(d,2np)$ reactions, will be also shown for the population of $^{54}$Mn residual nucleus.

Nevertheless, the transfer reactions are important at low incident energies, then decreasing with the deuteron energy increase, while the BU excitation function becomes continuously larger. 
Finally, consideration of the deuteron incident--flux decrease due to its absorption within BU as well as DR processes provides the correct total cross-section going towards PE+CN statistical decay of the excited system. Thus, a reduction factor of the total-reaction cross section due to the direct interactions (DI) of the breakup, stripping and pick-up processes accounted in the present analysis is given by:
\begin{eqnarray}\label{eq:2}
1 - \frac{\sigma_{BU} + \sigma_{(d,p)} + \sigma_{(d,t)}}{\sigma_R} %\: \nonumber \\
     = 1 - \frac{\sigma_{DI}}{\sigma_R}.
\end{eqnarray}
Its energy dependence is shown in Fig.~\ref{55Mnd_DIf} (c) at once with that corresponding to BU and DR reaction mechanisms, pointing out the role of each one in the deuteron interaction process with $^{55}$Mn target nucleus.
First, one may note the high importance of $\sigma_{DI}$/$\sigma_R$ at lowest incident energies due to the above-mentioned behavior of the stripping excitation function. The decrease of the DR component leads to a steep increase with the deuteron energy of the PE+CN weight, in spite of the BU increase. The PE+CN fraction maximum, around $\sim$ 64\% at energies of 10-12 MeV, is followed by a decrease due to the continuous increase of BU with the incident energy. Thus, both DI and PE+CN cross sections shown in Fig.~\ref{55Mnd_DIf}(c) have values close to half of $\sigma_{R}$ at energies around 42 MeV, then DI dominates with $E_d$ increase. It is thus pointed out the important role of the DI mechanisms for deuteron interactions. 

\subsection{Statistical emission}
\label{PE+CN}

The PE and CN statistical processes become important with the increase of the incident energy above the Coulomb barrier (e.g., Ref. \cite{RC2015}). The corresponding reaction cross sections have been calculated using the TALYS-1.9 code \cite{talys} and the reduction factor of Eq.~\ref{eq:2} in order to take into account the above-mentioned absorption of he deuteron flux into the DI processes.

The following input options  of the TALYS-1.9 code have been used: (a) the OMPs of Koning-Delaroche \cite{KD}, Daehnick {\it et al.} \cite{dah}, Becchetti-Greenlees \cite{BG}, and Avrigeanu {\it et al.} \cite{AHA} for neutrons and protons, deuterons, tritons, and $\alpha$-particles, respectively, (b) the back-shifted Fermi gas (BSFG) formula for the nuclear level density, (c) no TALYS breakup contribution, since the above-mentioned BF enhancements is still under implementation in TALYS, and (d) the PE transition rates calculated by means of the corresponding OMP parameters also involved within BU, DR, and CN calculations, for a consistent use of the same common parameters within various mechanism models. One may note that the same PE option was used with previous similar analyses of deuteron interaction with V, Cr, Fe, Co, Ni, and Cu \cite{VCo,Crd,Fed,Nid,Cud} leading finally to an improved agreement with the measured data. %, using the value 3 for the 'preeqmode' keyword.

\begin{figure*}
\resizebox{2.09\columnwidth}{!}{\includegraphics{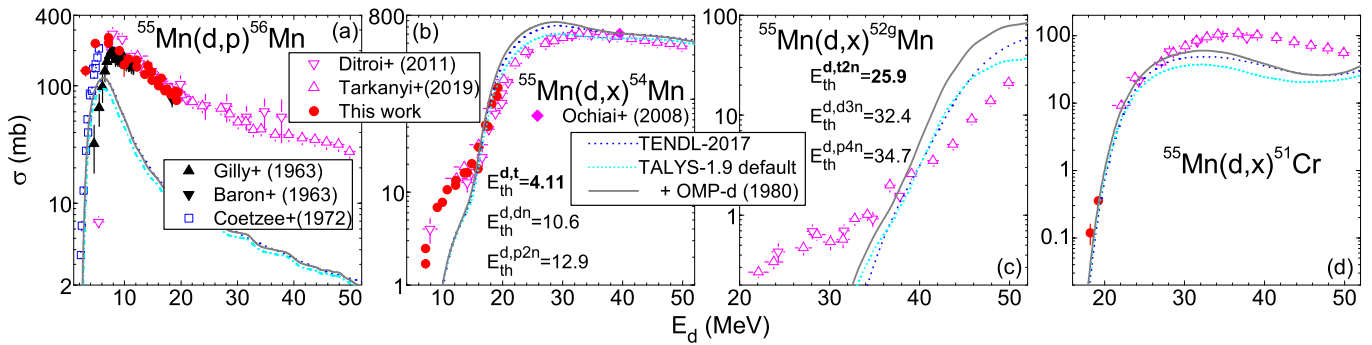}}
\caption{(Color online) Comparison of previous \cite{ditroi,tark,gil,baron,coet,ochi} and present (full circles) measured data, evaluated \cite{TENDL} (short-dashed curves), and calculated results obtained with TALYS-1.9 code \cite{talys} using either its whole default input (short-dotted) or the replacement of default deuteron OMP by that of Ref. \cite{dah} (solid curve), of deuteron-induced reactions on $^{55}$Mn.}
\label{55Mnd_def}
\end{figure*}
\begin{figure*}
\resizebox{2.09\columnwidth}{!}{\includegraphics{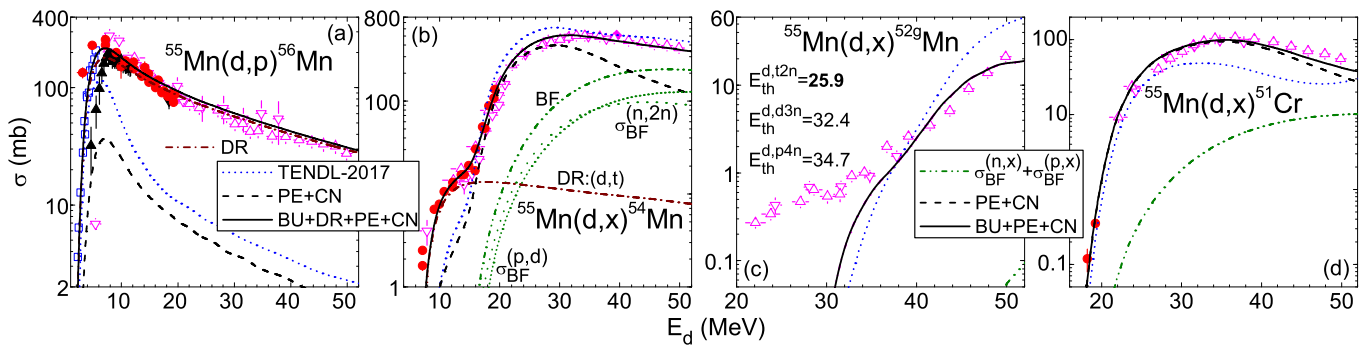}}
\caption{(Color online) As Fig.~\ref{55Mnd_def} but for present calculated results (solid curves) as well as for the PE+CN component (dashed curves), and (a) stripping $(d,p)$ and (b) pick-up $(d,t)$ (dash-dotted curves), BF enhancement (dash-dot-dotted curve) as sum of either $(n,2n)$ (dotted curve) and $(p,d)$ (short-dotted curve), or (c,d) the $(n,x)$ and $(p,x)$ reactions (see text).} 
\label{55Mnd_fin}
\end{figure*}

\section{RESULTS AND DISCUSSION}
\label{Activation}

The excitation functions of the residual nuclei $^{56,54}$Mn and $^{51}$Cr, measured in the present work for deuterons on $^{55}$Mn at energies $\leq$20 MeV (Sec.~\ref{exp}), are compared in Fig.~\ref{55Mnd_def} with the data formerly available up to 50 MeV \cite{ditroi,tark,gil,baron,coet,ochi}, the corresponding TENDL--2017 evaluation \cite{TENDL}, and results of calculation using TALYS-1.9 code and its default input parameters \cite{talys}. The new measured data of $^{54,56}$Mn activation are in agreement especially with the recent measurement of Tark\'anyi {\it et al.} \cite{tark}, extending to lowest energies the accurate experimental description. The same extension is provided by the new cross sections for $^{51}$Cr activation measured for the first time below 20 MeV.

The comparison with the most recent evaluated \cite{TENDL} and TALYS-1.9  \cite{talys} default calculation results proves however significant cross-section differences for all these reactions. A similar case is that of $^{52g}$Mn activation that was recently measured within a larger energy range (20--50 MeV) \cite{ditroi,tark} as shown in Fig.~\ref{55Mnd_def}(c). It has also been involved in this work for an overall analysis of the present model approach. 

A first effect taken into account to obtain a better measured-data account concerned the adopted input parameters of the deuteron OMP. Thus, taking the advantage of the elastic-scattering analysis given in Sec.~\ref{omp}, the TALYS default option of the deuteron double-folding OMP \cite{watomp} of nucleon global OMPs \cite{KD} was replaced with the OMP parameters of Daehnick et al. \cite{dah}. However, despite the substantial difference between the related elastic-scattering angular distributions shown in Fig.~\ref{55Mnd_elasticf}, the corresponding changes of the calculated cross sections are much smaller. There is particularly apparent (i) the continuing underestimation of the $(d,p)$ reaction data with even an order of magnitude, (ii) the failure in accounting the first decade above the effective thresholds of $^{54,52g}$Mn activation, and (iii) the good agreement in the same energy range but followed then by an underestimation with a factor $\sim$5 for $^{51}$Cr activation (Fig.~\ref{55Mnd_def}). So distinct discrepancies can be obviously related to the complexity of the interaction process, not entirely accounted for in routine evaluation/theoretical analyzes. 

The careful analysis of all involved reaction mechanisms as discussed in Sec.~\ref{models} may lead however to the well improved agreement of measured data and model calculation results, as shown in Fig.~\ref{55Mnd_fin}. 
The mechanism detailed contributions are particularly illustrated too, pointing out the strength of each one. 
Additional comments concern the reaction types and residual nuclei as follows. 

\subsection {The $^{55}$Mn$(d,p)^{56}$Mn reaction} 

The analysis of the population of $^{56}$Mn residual nucleus through deuteron interaction with $^{55}$Mn represents actually a distinct test of the reaction model approach due to the dominant contribution of the stripping DR mechanism. 
The comparative analysis of all reaction mechanism contributions involved in the $^{55}$Mn$(d,p)^{56}$Mn reaction, shown in Fig.~\ref{55Mnd_fin}(a), points out that the PE+CN component is lower by more than one order of magnitude than the stripping mechanism.% which is indeed essential for the suitable account of the measured excitation function. 
The inelastic breakup enhancement brought by breakup neutrons through $^{55}$Mn(n,$\gamma$)$^{56}$Mn reaction is practically negligible and not visible in this figure. 

In fact, the major underestimation of the experimental data by TENDL-2017 evaluation is just due to the overlooking of the key role of direct stripping process. Actually, this proof is just in line with the previous discussions of the $(d,p)$ excitation functions for $^{51}$V \cite{VCo}, $^{50}$Cr \cite{Crd},  $^{58}$Fe \cite{Fed}, $^{64}$Ni \cite{Nid}, and $^{93}$Nb \cite{Nbd} target nuclei, which show apparent discrepancies between the measured data and TENDL evaluations.

\subsection {The $^{55}$Mn$(d,x)^{54}$Mn reaction} 

The analysis of the $^{55}$Mn$(d,x)^{54}$Mn reaction [Fig.~\ref{55Mnd_fin} (b)] is the most interesting one from the viewpoint of the variety of contributing reaction mechanisms. 
This residual nucleus is populated entirely through the pick--up reaction $(d,t)$ at the incident energies lower than $\sim$11 MeV, i.e. below the thresholds for emission of additional particles in $(d,nd)$ and $(d,p2n)$ reactions. 
The DR component becomes then not significant, with a contribution which is with an order of magnitude lower than the dominant channels above 20 MeV. 
The latest are PE+CN that increase faster and reach a maximum around 29 MeV. 
Then, two inelastic-breakup enhancing contributions,  through the $^{55}$Mn(n,2n)$^{54}$Mn and $^{55}$Mn(p,d)$^{54}$Mn reactions induced by breakup-nucleons, become prevailing with the energy increase. 
Thus, they are $\sim$10\% of PE+CN contribution around 20 MeV incident energy, but over that above 42 MeV. This main reaction-channel interchange leads to a slower decrease of the excitation function comparing to its steep increase above the threshold. 

Altogether, the sum of the five reaction contributions succeeded to describe the measured $^{55}$Mn$(d,pxn)^{54}$Mn cross sections along the whole energy interval. Moreover, the underestimation of the lowest-energy data by the TENDL-2017 evaluation could be just the effect of the $(d,t)$ pick--up process overlooking as it was stressed out in previous analysis of deuteron interaction with $^{nat}$Ni \cite{Nid} and $^{93}$Nb \cite{Nbd}.

\subsection{The $^{55}$Mn$(d,x)^{51}$Cr reaction}

The new data measured in this work for $^{55}$Mn$(d,x)^{51}$Cr reaction are important for completing the $^{51}$Cr excitation function at incident energies just above its effective threshold, while previous measurements covered higher energies from 22 up to 50 MeV \cite{ditroi,tark} [Fig.~\ref{55Mnd_fin}(d)]. 
Among possible reaction channels, $^{55}$Mn$(d,\alpha 2n)^{51}$Cr represents the main contribution to the $^{51}$Cr population in the whole energy interval, as it is suggested by the shape of the excitation function.

As stressed out in previous analyses of deuterons interacting with $^{nat}$Fe \cite{Fed},  $^{nat}$Ni \cite{Nid}, and $^{nat}$Cr \cite{Crd}, the measured activation cross sections of $^{51}$Cr residual nucleus is a cumulative process (see Fig. 18 of Ref. \cite{Fed}), the $EC$ decay of $^{51}$Mn residual nucleus to $^{51}$Cr being much shorter ($T_{\frac{1}{2}}$=46.2 min) than the measurement time of the induced activity. 
However, because the population of $^{51}$Cr residual nucleus has been measured for incident energies up to 50 MeV, the contribution brought by $^{51}$Mn  decay should be considered negligible due to the high energy thresholds of the reactions populating it. These are 36.8 MeV for $^{55}$Mn$(d,t3n)$ and 45.6 MeV for $^{55}$Mn$(d,p5n)$. 

A suitable account of $^{51}$Cr excitation function has been obtained in this work taking into consideration the statistical PE+CN mechanisms as well as the inelastic-breakup enhancement brought by breakup-nucleons interaction with $^{55}$Mn. The latter contribution becomes significant above the incident energy of 40 MeV, improving well the agreement of the measured data and model calculation.

\begin{figure} [b]
\resizebox{0.8\columnwidth}{!}{\includegraphics{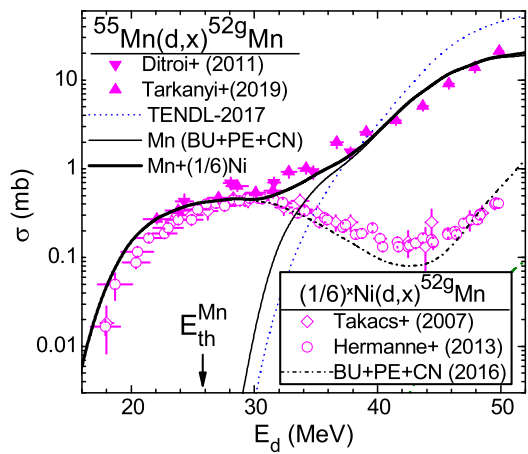}}
\caption{As Fig.~\ref{55Mnd_fin}(c) but with additional data \cite{tak07,herm} and similarly calculated results (short dash-dotted curve) for Ni$(d,x)^{52g}$Mn reaction, times a factor of 1/6, and the sum (thick solid curve) of these results and those for  $^{55}$Mn$(d,x)^{52g}$Mn reaction (thin solid curve).}
\label{55Mnd_normfin}
\end{figure}

\subsection{The $^{55}$Mn$(d,x)^{52g}$Mn reaction}

Despite the measured cross sections in this work up to only 20 MeV, a suitable account of the accurate recent measurements of $^{52g}$Mn activation between 20--50 MeV \cite{ditroi,tark} has been essential for a consistent and complete analysis of deuteron-activation of manganese. 
Actually, a reliable understanding of the variety of deuteron interaction processes should involve nevertheless the whole range of low incident energies, i.e. $\leq$50 MeV.

Population of $^{52g}$Mn residual nucleus follows mainly the PE+CN statistical mechanisms, as it results from Fig.~\ref{55Mnd_fin}(c). 
Thus, there is a BF enhancement due to breakup nucleons, that is weaker by at least two orders of magnitude. 
It becomes visible, in comparison with the measured data \cite{ditroi,tark}, only above 50 MeV.
However, a sudden underestimation of these data appears below $\sim$37 MeV. 
On the other hand, this underestimation exists below the thresholds of possible reactions populating $^{52}$Mn, as 25.9 MeV for $^{55}$Mn$(d,t2n)$, 32.4 MeV for $^{55}$Mn$(d,d3n)$, and  34.7 MeV for $^{55}$Mn$(d,p4n)$.

At this point, it became of interest to take into account the fact that these excitation functions  \cite{ditroi,tark} were measured using targets consisted of a natural high purity Ni(2\%)-Mn(12\%)-Cu(86\%) alloy. 
The authors considered indeed that in principle the investigated products could, apart from reactions on $^{55}$Mn, also be produced by nuclear reactions on the other two alloy components. 
Based on their published results for activation of these product nuclides by deuterons on Ni \cite{tak07,herm} and Cu, they found that only negligible corrections had to be introduced to their measured data derived for $^{55}$Mn. 

Nevertheless, we have included in Fig.~\ref{55Mnd_normfin} the measured data  and calculated cross sections for $^{52g}$Mn activation by deuterons on Mn [Fig.~\ref{55Mnd_fin}(c)] as well as on Ni \cite{tak07,herm} but reduced with a correction factor 1/6 corresponding to the Ni/Mn relative amount within the target alloy. 
The final agreement shown by this comparison supports indeed a Ni contribution within the measurements for Mn at the incident energies below the lowest threshold of the above-mentioned reactions on $^{55}$Mn. 
This contribution becomes then negligible at higher energies as well as in the whole energy range, as well as within two orders of magnitude for the other three reactions in Fig.~\ref{55Mnd_fin}.

On the other hand, the calculated excitation function of $^{nat}$Ni$(d,x)^{52g}$Mn \cite{Nid}, with the strongest contribution coming from $(d,2\alpha)$ reaction on $^{58}$Ni and threshold of 1.28 MeV, is different by that shown in Fig. 21 of Ref. \cite{Nid}. 
The previous one has concerned the total activation of $^{52}$Mn. 
Finally, the addition of this calculated contribution of Ni activation in the target alloy of Refs. \cite{ditroi,tark} to the model results shown formerly in Fig.~\ref{55Mnd_fin}(c) seems to describe well the measured data reported for $^{52g}$Mn activation by deuterons on Mn. It is thus confirmed the suitable account of all available measured cross sections for deuteron activation of Mn by the present model approach.

\section{CONCLUSIONS}
\label{Sum}

The activation cross sections for production of $^{54,56}$Mn, and $^{51}$Cr radioisotopes in deuteron-induced reactions on $^{55}$Mn are measured at incident energies up to 20 MeV. 
They are in good agreement with the previously reported experiments \cite{gil,baron,coet,ochi,ditroi,tark} while all of them are the object of an extended analysis from elastic scattering until the evaporation from fully equilibrating compound system. 

A particular attention has been given at the same time to breakup and direct reactions mechanisms. 
The mark BU rather than BF of breakup fusion, for the sum of various contributions to an activation cross section in Fig.~\ref{55Mnd_fin}, underlines the consideration of both breakup effects, i.e., the overall decrease of $\sigma_R$ due to incident flux leakage toward breakup, as well as the BF enhancement. 
A detailed theoretical treatment of each reaction mechanism has made possible a reliable understanding of the interaction process as well as accurate values of the calculated deuteron activation cross sections. 

Furthermore, the comparison of the experimental deuteron activation cross sections with our model calculations as well as the corresponding TENDL-2017 evaluation supports the detailed theoretical treatment of deuteron interactions.
The discrepancies between the measured data and that evaluation have been the result of overlooking the inelastic breakup enhancement and less appropriate treatment of stripping and pick-up processes. 
This comparison particularly indicates the importance of the new measured cross sections around the maximum of the $^{55}$Mn$(d,p)^{56}$Mn excitation function as well as the role of stripping mechanism to provide the suitable description of these data. 

The measured cross sections for $^{55}$Mn$(d,x)^{54}$Mn reaction at low incident energies in the present work play also a similar role, revealing the importance of the pick-up mechanism for the description of data around the reaction threshold. 
Actually, the main reaction-channel interchange in the energy range up to 50 MeV for this activation product underlines the complexity of the deuteron-induced reactions and need of a complete theoretical approach. 
The first measurement of $^{55}$Mn$(d,x)^{51}$Cr cross sections around the threshold has also been as effective for a suitable description of the whole excitation function as taking into account the BF above 40 MeV.

The overall agreement between the measured data and model calculations supports the fact that major discrepancies shown by the current evaluations are due to missing the proper account of direct interactions. The consistent theoretical frame of the deuteron interactions supported by advanced codes associated to the nuclear reactions mechanisms provides predictability in addition to the use of various-order genuine Pad\'e approximations \cite{hermanne,engle} needed in applications.

However, while the associated theoretical models for stripping, pick-up, PE and CN are already settled, an increased attention should be paid to the theoretical description of the breakup mechanism including its inelastic component. The recently increased interest on the theoretical analysis of the breakup components (e.g., \cite{neoh,carlson,CDCC4,lei15,lei18}) may lead eventually to the refinement of the deuteron breakup empirical parametrization and increased accuracy of the deuteron activation cross section calculations.

Nevertheless, improvement of the deuteron breakup description requires, apart from the increase of its own data basis, also complementary measurements of $(d,px)$ and $(n,x)$, as well as $(d,nx)$ and $(p,x)$ reaction cross sections for the same target nucleus, within corresponding incident-energy ranges. Moreover, as it has been proved in this work, additional measurements of deuteron-induced reaction cross sections especially just above reaction thresholds but also close to 50 MeV would provide further opportunities to get a better understanding of these complex reactions.

\section*{Acknowledgments}
This work has been partly supported by OP RDE, MEYS, Czech Republic under the project SPIRAL2-CZ, CZ.02.1.01/0.0/0.0/16\_013/0001679 and by Autoritatea Nationala pentru Cercetare Stiintifica (Project PN-19060102), and carried out within the framework of the EUROfusion Consortium and has received funding from the Euratom research and training programme 2014-2018 and 2019-2020 under Grant Agreement No. 633053. The views and  opinions expressed herein do not necessarily reflect those of the European Commission.

\end{document}